# Municipal cyber risk modeling using cryptographic computing to inform cyber policymaking

*A case study in the policy contributions of cyber risk measurement for municipal cybersecurity[1]*

**Authors[2]:** Avital Baral, Taylor Reynolds, Lawrence Susskind, Daniel J. Weitzner, Angelina Wu

The authors would like to thank Rebecca Spiewak for her research building MIT's Ransomware Readiness Index that was used as the control list for the study and to Scott Shackelford and Etyan Tepper for their comments and suggestions as discussants at the CLPSC.

<div style="text-align:center; color:red">

Working Draft for Presentation at the
Cybersecurity Law and Policy Scholars Conference
September 29, 2023

</div>

## Abstract

Municipalities are vulnerable to cyberattacks with devastating consequences, but they lack key information to evaluate their own risk and compare their security posture to peers. Using data from 83 municipalities collected via a cryptographically secure computation platform about their security posture, incidents, security control failures, and losses, we build data-driven cyber risk models and cyber security benchmarks for municipalities. We produce benchmarks of the security posture in a sector, the frequency of cyber incidents, forecasted annual losses for organizations based on their defensive posture, and a weighting of cyber controls based on their individual failure rates and associated losses. Combined, these four items can help guide cyber policymaking by quantifying the cyber risk in a sector, identifying gaps that need to be addressed, prioritizing policy interventions, and tracking progress of those interventions over time. In the case of the municipalities, these newly derived risk measures highlight the need for continuous measured improvement of cybersecurity readiness, show clear areas of weakness and strength, and provide governments with some early targets for policy focus such as security education, incident response, and focusing efforts first on municipalities at the lowest security levels that have the highest risk reduction per security dollar invested.

---

[1] The authors would like to thank Rebecca Spiewak for her research designing MIT's Ransomware Readiness Index that was used as the control list for the study. We appreciate the collaboration with Geoffrey Beckwith, Stan Corcoran, and Lin Chabra at the Massachusetts Municipal Association, Joe Callahan at Cabot Risk Strategies, as well as the municipal government cybersecurity professionals who assisted with data collection.

[2] Authors listed in alphabetical order. Weitzner and Wu were supported, in part, by NSF grant Collaborative Research: DASS: Legally Accountable Cryptographic Computing Systems (LAChS) Award Number: 21315415. Reynolds was supported by MIT's Future of Data Initiative, MIT's FinTech@CSAIL, and MIT's Machine Learning Applications @CSAIL.

## Table of contents

## Introduction

Municipalities are vulnerable to cyber attacks with devastating consequences. In the United States, local governments are also often in charge of critical infrastructure and key services for constituents. Additionally, they are commonly strapped for resources with limits on time, money, and expertise, particularly in the IT domain (Norris et al., 2021). This combination makes municipalities a frequent target of cyber attacks, including notable recent examples in Baltimore (Baltimore Sun, 2019a), Atlanta (Reuters, 2018) and a 2019 attack in Louisiana that forced the governor to issue a statewide emergency declaration (State of Louisiana, 2019). Cyber attacks on public infrastructure have concrete negative impacts on communities, from slowing down or stopping important municipality-mediated endeavors such as home sales paperwork (Baltimore Sun, 2019b), as well as significant impacts on individual lives (Spiewak, 2022). The US White House has convened a series of summits on issues such as strengthening the cybersecurity of US schools as a result (US White House, 2023).

Municipalities are reluctant to share information regarding their cybersecurity challenges. They worry that sharing data regarding the incidents they face will result in public backlash as well as open the door to further attacks (Preis and Susskind, 2022). In particular, while there exist surveys regarding the threats that municipalities (Norris et al., 2021), as well as organizations in other sectors face (US White House, 2018), there is a dearth of data regarding the frequency of attacks and losses sustained (de Castro et al., 2020). Given this lack of data and the decentralized nature of US local government, municipalities lack a way of benchmarking their preparedness against peers and miss the opportunity to learn from each other's experiences.

State and federal governments want to support municipal cyber security preparedness but lack adequate guidance on what kind of support will be most useful. Municipalities look to state and federal government for guidance and help with their cybersecurity challenges, but the lack of reliable data regarding the kinds of cyber threats, attacks, and losses that entities are facing limits the efficacy and adequacy of the help that those higher-level governments can provide. States have piloted different cybersecurity initiatives over the last decade aimed at municipalities such as regional CISO as a service programs (StateTech, 2020), Cyber Volunteer Corps (NGA, 2017), and employee trainings (Commonwealth of MA, 2023). However, a critical problem with these initiatives is that they are not being evaluated to test the effectiveness in reducing cyber attacks against municipalities. This is because there is very little information about attacks and losses that is available. Organizations are reluctant to share information about current security practices, details about security control failures, and the scale of losses. States are well-intentioned in their desire to help protect municipalities, but without robust data, it is challenging to evaluate the usefulness of different interventions, which ones to keep and expand, and which ones are a poor allocation of limited resources. There are, however, new methods for securely and privately collecting cyber risk data that can overcome these challenges.

Cryptographic techniques such as secure multiparty computation can help collect sensitive data without requiring supplying parties to identify themselves or disclose their data. We demonstrate that if municipalities would leverage these tools we describe to build up metrics and models in evaluating their cyber risks, then municipalities and policymakers at the state and national level would be able evaluate and prioritize cyber hygiene measures, and thereby minimize risks overall. This could take the form of a set of "reasonable" cybersecurity measures for municipalities. The Secure Cyber Risk Aggregation and Measurement (SCRAM) system (de Castro et al., 2020) is a platform that uses secure multiparty computation to aggregate participant responses to customizable questionnaires regarding cybersecurity controls implementation, cybersecurity attacks, and related monetary and institutional losses (Figure 1). Participants in SCRAM computations receive aggregate scores for the sector for metrics laid out by the questionnaire without being required to disclose their own data to any third party. That means that neither the participants nor the computation administrators can access any participants' responses. The outcome is more robust data regarding cybersecurity-related attacks and losses in which participants are incentivized to honestly participate in given the privacy of the process. The SCRAM platform has been used in pilot computations with large firms across different sectors.

Figure 1: **SCRAM process for gathering data using secure multi-party computation**

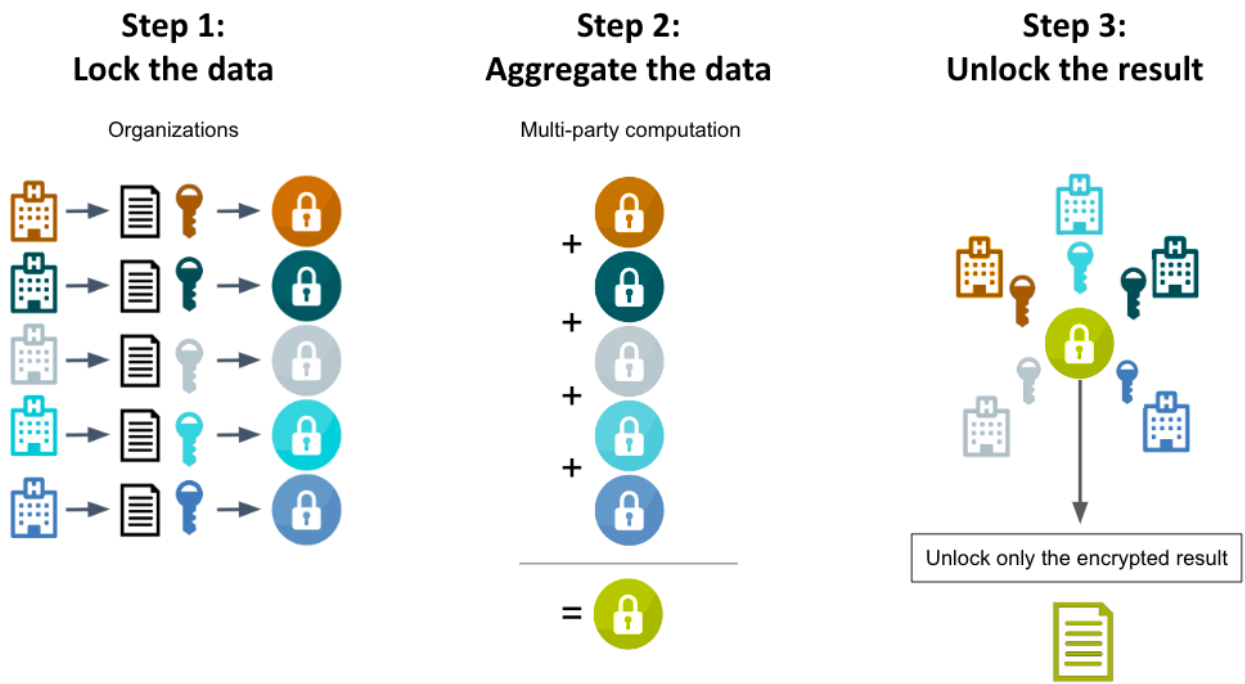

This paper provides a proof of concept towards providing vital insight into the municipal cybersecurity data problem using data from a collection across municipalities in a single US state. It shows the types of data that will be available to cyber policymakers in the future. We adopt an existing set of questions from the MIT Ransomware Readiness Index (Spiewak et al., 2021) and then collect the data in SCRAM to build the benchmarks, derive the indicators, and produce risk models and forecasts. These new data can then inform policy decisions at all levels of government.

There are several key outputs from our research:

1. **Benchmarks:** Security benchmarks that organizations can use to compare their own security posture against their peers
2. **Risk forecasts:** Risk forecasts can provide estimates of the risk for an organization, an indication of how financial risk changes with varying maturity levels of controls to prioritize security investments, and information on a fair price for insurance coverage.
3. **Policy considerations**: Benchmarks for the group provide important insights into how the municipalities perceive the maturity level of their own security controls. Policymakers can use these insights to target training, funding, and interventions to improve the overall security of critical infrastructure.

## Literature review

There are a handful of national surveys regarding cybersecurity in the public sector and together they paint a dire picture of state and municipalities' cybersecurity preparedness. Municipalities report themselves to be unprepared to effectively face the threat of cyberattack, chiefly citing lack of funding, lack of qualified staff, and lack of prioritization of information technology by their organizational leadership as barriers to an adequate cybersecurity posture. Norris et al (2019) conducts a nationwide survey of municipalities and finds that only a minority of responding municipal IT professionals considered their municipalities well-prepared to detect and recover from ransomware attacks and cyberattacks generally (Norris et al., 2019). Earlier work by Norris identifies top barriers to cybersecurity readiness in municipalities as the inability to pay competitive salaries to IT professionals, lack of funding for cybersecurity, and lack of dedicated security staff (Norris et al., 2019). A consistent and concerning thread through the surveys is that a significant portion of respondents report that they do not know if or how often they are being attacked and are unable to rate their own levels of preparedness, indicating a lack of preparation and prioritization of cybersecurity (Norris et al., 2019); (Preis and Susskind, 2022). Among those respondents that did report attack attempt frequencies, 27.7% of respondents reported being attacked at least hourly, and another 19.4% reported being attacked at least once a day (Norris et al., 2019).

A biennial survey of state CIOs conducted jointly by Deloitte and the National Association of State CIOs (NASCIO) repeatedly affirms the findings above (Deloitte and NASCIO, 2018; Deloitte and NASICO, 2020). The Deloitte-NASCIO reports highlight the need for greater cybersecurity funding at the state level and for a coherent, centralized policy strategy for local government. CISOs consulted in the 2018 and 2020 Deloitte-NASCIO surveys reported that they viewed state-level regulations backed by funding commitments as most effective in improving cybersecurity posture. Only a minority of states had a cybersecurity budget line item (Deloitte and NASCIO, 2018; Deloitte and NASICO, 2020)

A 2020 survey of the Coalition of City CISOs in partnership with the International City/County Managers Association (ICMA) of chief security officers serving 14 large American cities echoed the findings of prior surveys and found that attacker demands for ransoms had increased in 2020 (Norris, 2021). The ICMA survey identified leadership buy-in as another top barrier to improving cybersecurity posture in local governments.

The state of Indiana produced a "State of Hoosier Cybersecurity" in 2020 that ran a large-scale survey of 178 organizations in Indiana (Boustead and Shackelford, 2020). The report found 19% of respondents had encountered a "cyber incident" in the previous three years. The report also found that half of the organizations indicated that they had cyber insurance (Boustead and Shackelford, 2020).

Municipalities are hesitant to share information about their cybersecurity posture and incidents they have faced due to concerns about further opening themselves up to cyberattack as well as political fallout (Preis and Susskind, 2022). The 2018 White House Council of Economic Advisors report, The Cost of Malicious Cyber Activity to the U.S. Economy (US White House, 2018), found an extremely broad range for the cost of cyberattacks directed at state and local governments, chronicling successful breaches with costs from USD 665,000 to USD 40.53 million, with a median cost varying from USD 60,000 to as high as USD 1.87 million.

There is additional money that will be allocated at the federal level in the United States to help local governments with cybersecurity. In 2023, DHS announced USD 375 million in funding to boost state and local cybersecurity preparedness (US DHS, 2023). While the allocation is large in the aggregate, it would amount to an average of just over USD 4,100 for each local government once spread across the 90,875 local government units identified by the US Census in 2022 (US Census, 2023).

The US Cybersecurity and Infrastructure Security Agency (CISA) will also begin collecting data on ransomware attacks when an organization reports their incident to the Federal Bureau of Investigation (FBI), CISA, or the US Secret Service (US CISA, 2023). It is unclear at this time whether the data compiled from each of the reporting agencies will be comparable.

While data on municipal cybersecurity is relatively new, there is a much longer history of cyber risk approaches. There are some well-known approaches for modeling cyber risk, but most of them are focused on individual risks and decisions related to specific control applications rather than a broader, holistic view of cyber risk for an organization.

Early approaches introduced a simple way of evaluating cyber risk. In the 1970s, Courtney posited that risk to electronic data processing systems could be summarized with two elements (Courtney, 1977):

1. Statement of impact - How badly a specific difficultly hurts
2. Statement of probability of encountering that difficult in a specified period of time

Nearly five decades later, Courtney's initial approach has largely held. The popular cyber risk modeling framework Factor Analysis of Information Risk (FAIR) has the same top-level structure and provides a framework for coming up with top-level estimates (Freund and Jones, 2014).

FAIR is commonly used in large sectors such as financial services. Both Courtney and FAIR's approach disseminate all risk measures from two components: frequency and impact. In FAIR, the implementation of security controls feeds into the frequency side, but at a low level in the structure.

Security and risk teams specifically focus on the impact of controls on risk outcomes, so other modeling approaches pull up cyber defenses to a top line element in the risk model alongside frequency and impact. For example, the CRAM model from Mukhopadhyay introduces a variable called "$Sec_T$" for security technology to capture control decisions at a high level (Mukhopadhyay et al., 2019).

In summary, the current literature shows low levels of security maturity in municipalities and a hesitation to share data about successful attacks. There are programs in place to support municipal cybersecurity across the various levels of government, but currently there is no way to evaluate the effectiveness of these programs. The basic cyber risk structures are well known however, and they can be applied to the municipal case using specific data that can be collected via security computations.

## Data

The data were collected in June 2022 from the municipalities in Massachusetts via a secure questionnaire prepared for the SCRAM platform. A description of each of the data elements is provided below. It is important to note, however, that the results from a multi-party computation are only aggregate figures and cannot be linked back to an individual organization that submitted data. The increase in privacy and security comes at the expense of no longer being able to see the individual inputs into a data aggregation.

All data are self-reported by municipalities, and participants are trained during a session on how to fill out the forms and estimate the data based on their own experiences. This training step is necessary to improve the comparability of the responses across firms.

The collected data that are used in this analysis include:

- **Maturity level:** 22 questions on the maturity level of controls (see Annex 1)
- **Incident count:** 1 question on the frequency of incidents
- **Control failures:** Count of the times individual controls failed leading to incidents
- **Financial costs (total):** 1 question on the total cost of incidents
- **Financial costs (control failures):** Data on the attributed costs of incident failures to specific controls

## Model

In our model, the annual cyber risk for a firm is determined by the historical frequency of attacks in the sector multiplied by a defense gap index scalar and by the average impact of incidents for the sector (Equation 1). Our key modeling assumptions for the municipal risk analysis are detailed in Annex 2.

(Eq 1) $$Frequency * DefenseGapIndex * Impact = AnnualRisk$$

Frequency is derived from data on incident counts in the data collection while impact data come from self-reported losses from those incidents. The Defense Gap Index is more complicated to build, but essentially provides a multiplier for risk measures that is based on how far the security of an organization deviates from the average security posture of its peers. Organizations with lower security levels have higher values with higher associated risks, while those with better security have reduced risk values that emerge. The Defense Gap Index also assigns more weight to control failures with a history of attributed losses, so organizations with lower security in these areas are penalized more than for a similar gap in a controls that has no losses associated with it. Finally, the Defense Gap Index is mapped to a loss distribution that emerges from the secure computations where firms with lower security have higher losses than those with better security. The details of the data and models are provided in Annex 2.

## Results

This section begins by providing the benchmarking results from the municipal computation related to control maturity, frequency of control failures, and corresponding financial losses. Next, the section takes these benchmarks and uses them to develop risk models for the group.

The data are all self-reported, which introduces biases and measurement challenges, but it still offers the best insight we have into the current state of municipal cybersecurity defenses in a state. We have worked to reduce these biases by holding multiple training sessions with the respondents to harmonize expectations, measurement parameters, and answer questions they have.

## Benchmarking

There are 83 municipalities from the same state that participated in the computation. Over a period of two years, the group reported 4 significant incidents and 14 control failures related to those incidents that were responsible for the losses.

The number of incidents across the group is low relative to the size of the group. From a security perspective, this is a desired outcome, but it does make the weights assigned to specific control failures and resulting losses potentially overweighted. As a result, researchers should keep these findings in context since small changes in the answers can have large effects when extrapolated out to the entire group. At the same time, these data can still provide us with preliminary information about the overall rates of incidents and losses, as well as the perception of control maturity across municipalities.

We run a total of six computations. The first includes all municipalities (n=83) and then the same metrics are computed for five subgroups based on population. All municipalities are weighted equally within each of the computation groups, so a larger city will have the same weight on the average as a smaller town in the same group (Table 2).

**Table 2: Classification criteria of each of the six computations**

| Computation 1 | All municipalities combined | n = 83 |
|---|---|---|
| Computation 2 | Municipalities with populations over 25,000 | n = 8 |
| Computation 3 | Municipalities with populations between 15,000 and 25,000 | n = 9 |
| Computation 4 | Municipalities with populations between 5,000 and 15,000 | n = 29 |
| Computation 5 | Municipalities with populations less than 5,000 | n = 16 |
| Computation 6 | Municipalities with no reported population* | n = 21 |

Note: Municipalities with no population can include school and fire districts that have no standard population but still participate in insurance pools.

The data collection does not separate out much larger cities, and if a significant number were included, they could affect the averages in different ways. But these results are applicable for somewhat smaller population groups of municipalities.

In terms of overall security levels, we do not find large variations between the different population sizes except for the smallest population groups (Figure 2). Unsurprisingly, we see that the largest municipalities report slightly higher maturity levels of their controls, while municipalities with smaller populations have less mature levels. One group of municipal entities does not include data on population and can include fire districts or school district in rural areas. These municipal entities with no population recorded had the lowest maturity levels in the overall group.

**Figure 2: Average maturity of controls, overall and by population size**

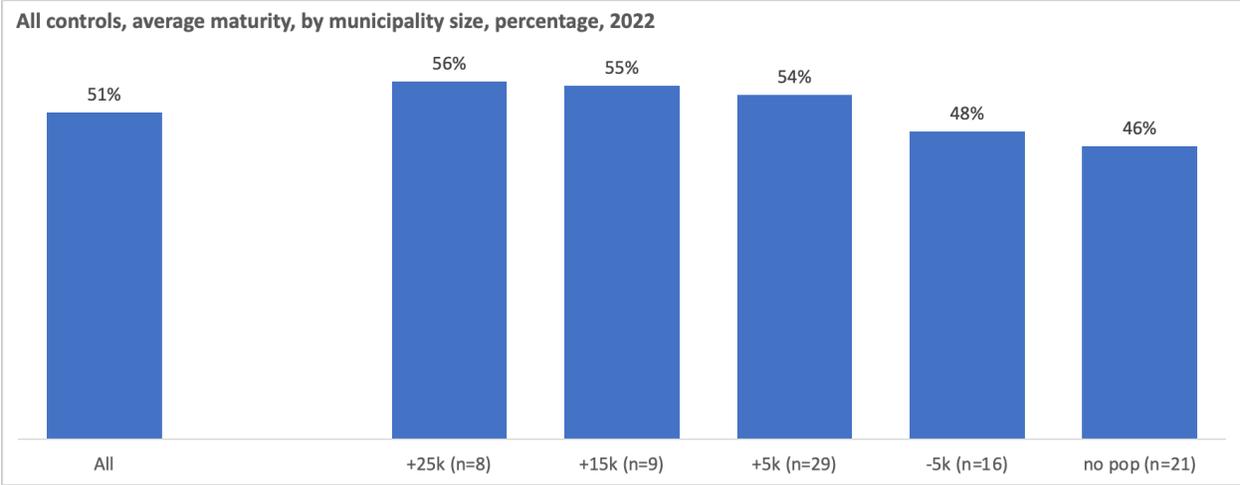

The average adoption rate across all individual controls was 51%, and that corresponds to essentially halfway between "partially implemented" and "largely implemented" on our scale (Table 1). It's important to note that the overall maturity rate of 51% is noticeably lower than the 65% maturity rate we have found among private firms in other sectors in forthcoming research. Even the largest subgroup of municipalities (with populations over 25k) had a lower maturity rate of 56% that is below the averages we find in private sector computations.

**Table 1: Implementation levels mapped to percentages for quantification**

| Not implemented | Partially implemented | Largely implemented | Fully implemented |
|---|---|---|---|
| 0% | 33% | 67% | 100% |

The 22 subcontrols in MIT's Ransomware Readiness Index are broken into 10 broader groups.[3] There are large differences in maturity levels across the 10 control groups for the 83 municipal entities (Figure 3).

**Figure 3: Security control maturity by category, average across municipalities**

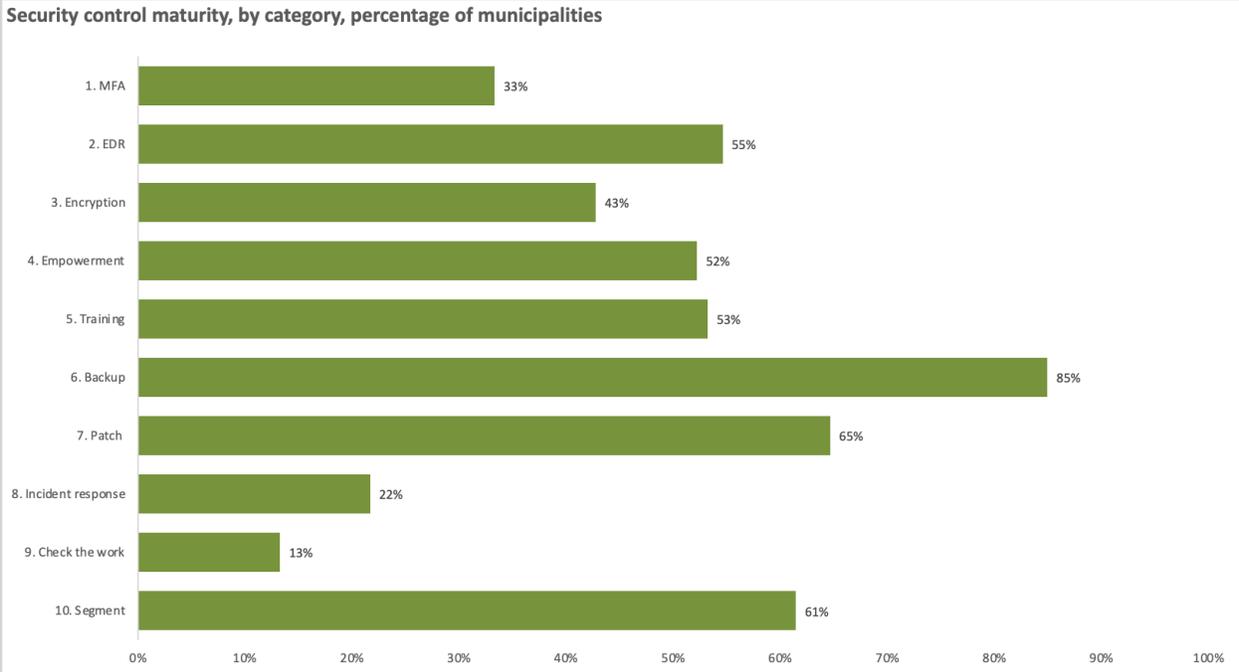

Security control maturity, by category, percentage of municipalities

| Category | Value |
|---|---|
| 1. MFA | 33% |
| 2. EDR | 55% |
| 3. Encryption | 43% |
| 4. Empowerment | 52% |
| 5. Training | 53% |
| 6. Backup | 85% |
| 7. Patch | 65% |
| 8. Incident response | 22% |
| 9. Check the work | 13% |
| 10. Segment | 61% |

## Areas of strength

The areas of highest perceived strength are backups, patching of systems, and segmenting the network. The backup category has the highest self-reported maturity of 85%. Within the backup group which scores highest across all categories, some elements have better maturity than others (Figure 4). The self-described maturity rating for performing backups on a regular basis is 95% (essentially fully implemented). However, the rating falls to 81% for testing backup data and 79% for storing backup data in an offline location – both of which are crucial for remediation efforts. This may imply that simply performing backups could be giving municipalities a false sense of security if those backups are not tested nor stored offline.

---

[3] Details of MIT's Ransomware Readiness Index are provided in Annex 1.

**Figure 4: Security control maturity, by control, average across all municipalities**

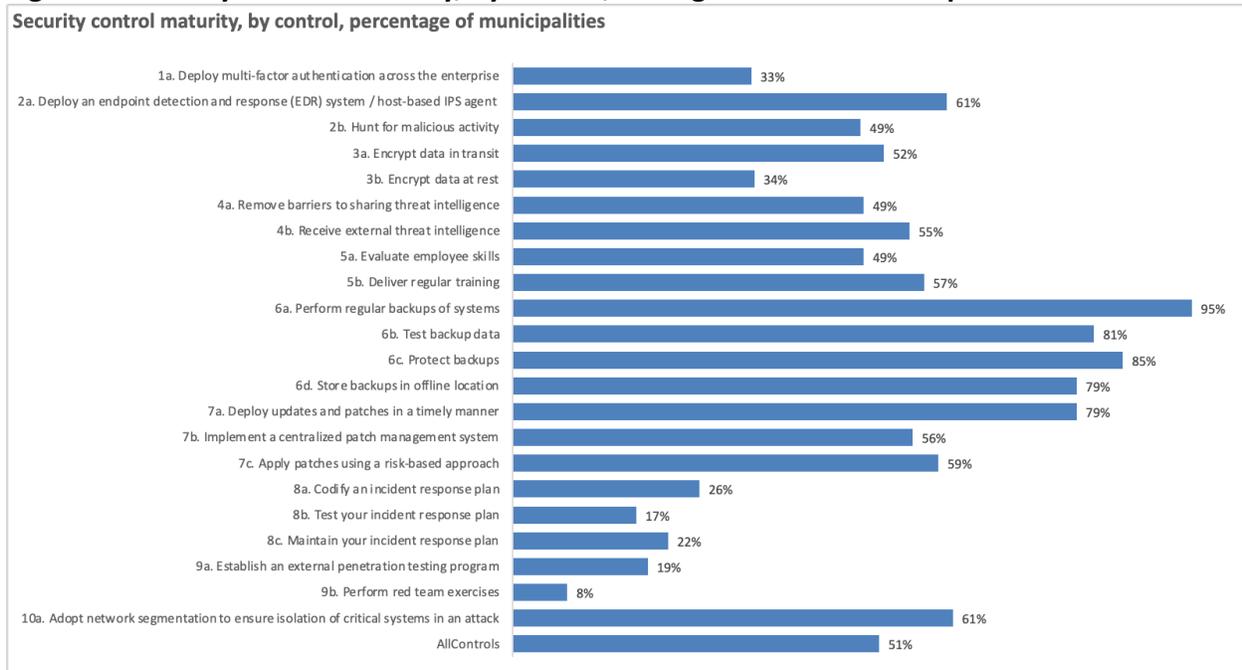

Security control maturity, by control, percentage of municipalities

| Control | Percentage |
|---|---|
| 1a. Deploy multi-factor authentication across the enterprise | 33% |
| 2a. Deploy an endpoint detection and response (EDR) system / host-based IPS agent | 61% |
| 2b. Hunt for malicious activity | 49% |
| 3a. Encrypt data in transit | 52% |
| 3b. Encrypt data at rest | 34% |
| 4a. Remove barriers to sharing threat intelligence | 49% |
| 4b. Receive external threat intelligence | 55% |
| 5a. Evaluate employee skills | 49% |
| 5b. Deliver regular training | 57% |
| 6a. Perform regular backups of systems | 95% |
| 6b. Test backup data | 81% |
| 6c. Protect backups | 85% |
| 6d. Store backups in offline location | 79% |
| 7a. Deploy updates and patches in a timely manner | 79% |
| 7b. Implement a centralized patch management system | 56% |
| 7c. Apply patches using a risk-based approach | 59% |
| 8a. Codify an incident response plan | 26% |
| 8b. Test your incident response plan | 17% |
| 8c. Maintain your incident response plan | 22% |
| 9a. Establish an external penetration testing program | 19% |
| 9b. Perform red team exercises | 8% |
| 10a. Adopt network segmentation to ensure isolation of critical systems in an attack | 61% |
| AllControls | 51% |

## Areas of weakness

Somewhat surprisingly, multi-factor authentication (MFA) has a relatively low maturity result of 33%, which is equivalent to partially implemented in our scale (Table 1). The lack of MFA is often considered a "deal killer" by the technical groups supporting underwriting for cyber insurance coverage, so organizations without MFA would struggle to find coverage from most insurers.

The breakdown of the MFA data by population size reveals some interesting insights. The lowest reported maturity levels come from mid-size municipalities, not the larger or smaller ones (Figure 5). This result is seen across several of the variables and could be explained by the dynamics of operations for large and small municipalities. The largest municipalities often have budgets to support dedicated IT staff that have some background in cybersecurity. The smallest set of municipalities often lacks a dedicated full-time IT staff member, and often outsources its IT services to firms with a specialization in the area. It could be precisely the mid-size municipalities that are large enough to have an on-staff IT member, but too small to have cyber specialists. We cannot know for sure without deeper investigation, but it is in an area for future exploration.

**Figure 5: Multi-factor authentication maturity, by municipality size**

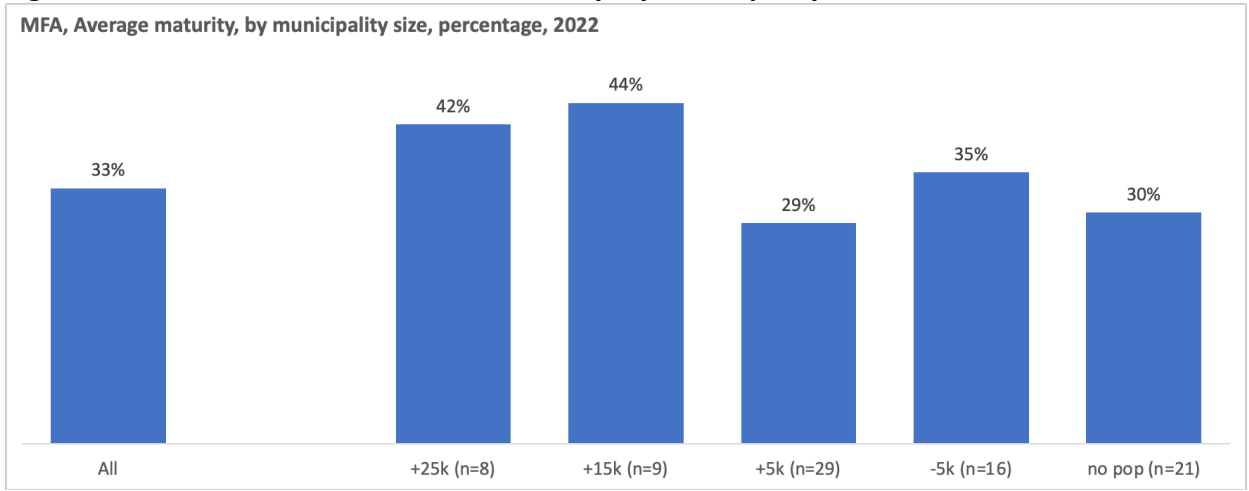

Data at a more-granular level on MFA responses highlights a particular challenge for municipalities (Figure 6). Only 4 out of 83 municipalities report fully implementing MFA, and 23 municipalities say that MFA is not implemented at all, even for very sensitive systems. Certainly, getting MFA implemented on these unprotected systems should be a key priority.

**Figure 6: MFA maturity counts, by maturity level, all municipalities, counts**

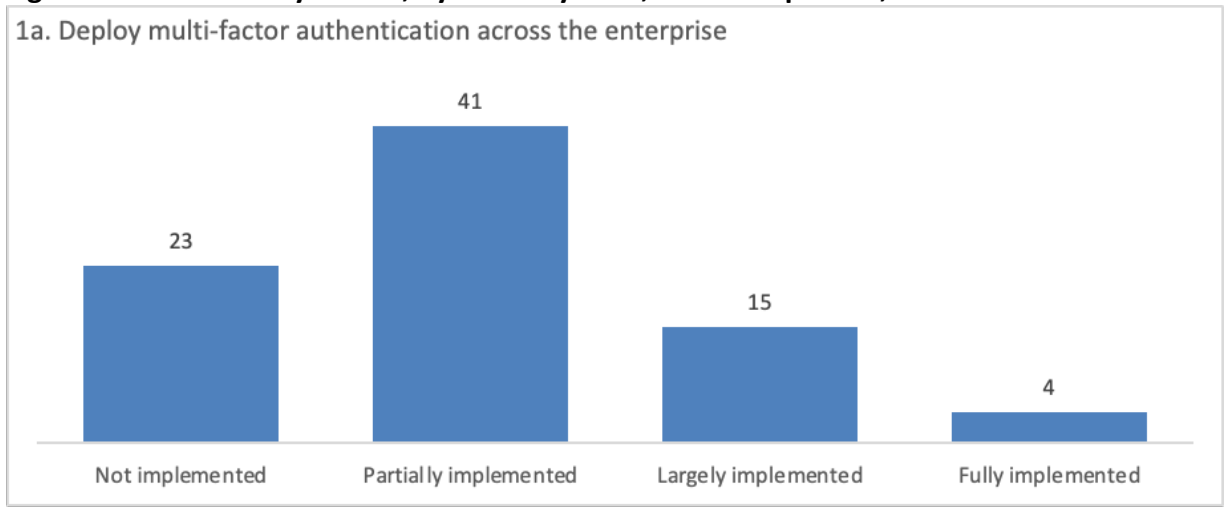

Encryption related questions reveal relatively low maturity levels. The average maturity rating for encrypting data in transit is 52%, which is roughly equivalent to the average maturity level for all controls reported by all municipalities. We would expect that these numbers would be high due to the widespread adoption of TLS and VPNs for connecting systems over a wide area network. On the other hand, encryption of data at rest is lower at 34%, which corresponds to "partially implemented" on our scale. In today's current threat environment, ransomware attacks offer a one-two punch of first exfiltrating data that can be released if ransoms are not paid and then encrypting data on the organization's own systems to extract payments. Encrypting data at rest, particularly sensitive data such as personally identifiable information (PII), is an important way to limit the impact of these attacks with an exfiltration component, particularly if the organization is relying on backups to recover and avoid paying ransoms. The low maturity score of 34% highlights that there is significant work to be done in this area.

The security controls with the lowest reported maturity levels were related to incident response and checking the work. The maturity levels across these five controls run from 8% to 26% and are significantly lower than the average across all controls of 51% (Figure 7).

**Figure 7:  Security controls with low rated maturity levels**

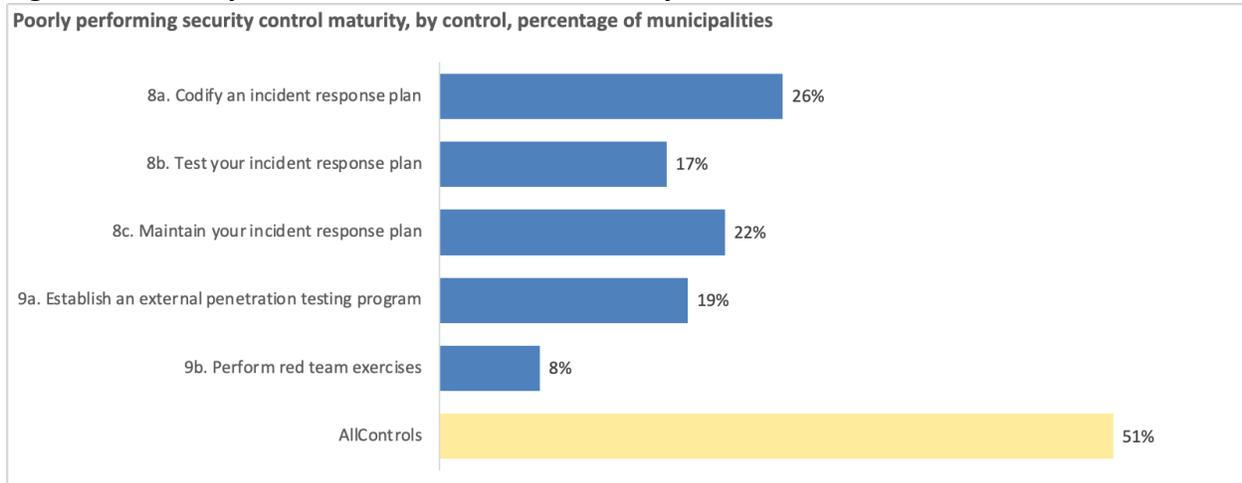

Municipalities largely lack outside testing of their security via red-team exercises or external penetration testing programs. It is not surprising that smaller communities may lack the resources to hire outside consultants to run these exercises.  A detailed look at the count of the responses for establishing an external penetration testing program shows that most municipalities have not implemented external penetration testing, but there are 12 municipalities that have either largely or fully implemented an external pen-testing process (Figure 8). Interestingly, informal conversations with participants indicate that the municipalities that do have fully or largely implemented red-teaming exercises may include some smaller municipalities that outsource their IT and security to specialized firms that run pen-testing programs over their portfolio of clients.

**Figure 8: Summary of responses for establishing an external pen-testing program, counts**

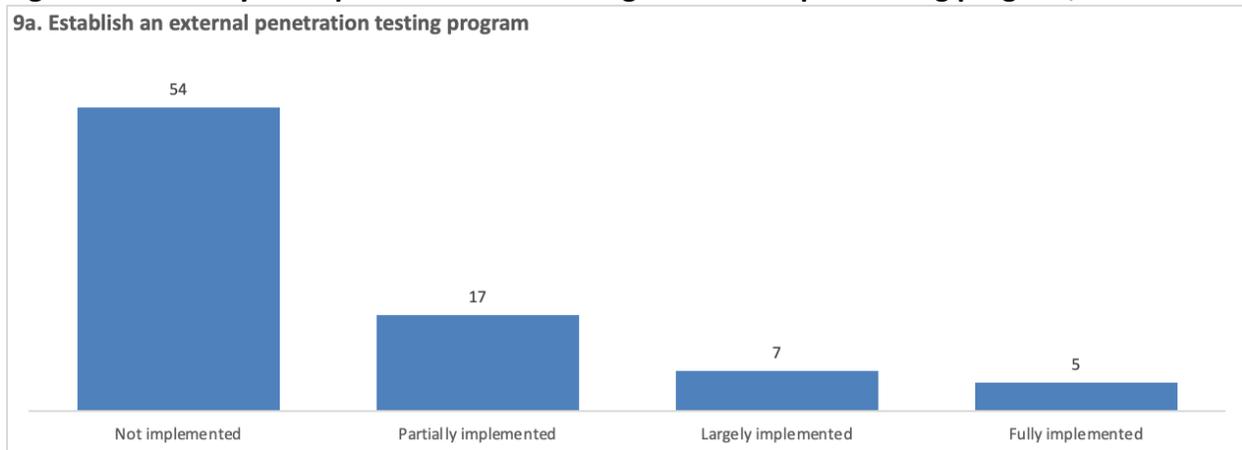

One of the more important findings is that municipalities rate their incident response maturity as very low even though all have access to state resources to develop incident response plans, and the insurer provides extensive incident response support if the municipalities ask for it. In this case, there seems to be a combination of a lack of awareness and missing out on resources at the state level that are already available to them.

*Benchmarking: Losses and control failures*

There are 4 significant incidents across the group of 83 municipalities over a period of 3 years. The municipalities that have incidents are asked to report up to 5 control failures that led to each incident among the 22 controls in the questionnaire. The total losses from the incident are then spread across the implicated controls equally. We give them equal weights because the attribution process is not precise enough currently to warrant an additional set of weights on responsibility for the incident.

The total losses across the four incidents are USD 628,000 which works out to an average of USD 157,052 for each incident. The SCRAM platform never has access to the raw inputs, so researchers do not have access to the magnitudes of any specific incident. However, by looking at the count of incidents in broad ranges, we can see that three of the losses were lower than US 100,000 and one loss was USD 500,000 or greater. We use the average and the general distribution of losses to model the impact of varying security controls on overall loss outcomes. Next, we examine the frequency of control failures and the losses attributed to them.

Figure 9 shows the number of failures attributed for each of the controls across the 4 incidents. Interestingly, the control for "deliver regular training" was implicated in 3 of the 4 incidents, and the "evaluate employee skills" was implicated in 2 of the 4. This highlights the difficulty of training users to avoid behaviors such as clicking on malicious links or opening infected attachments that can provide an entry point for attackers.

**Figure 9: Frequency of control failures, by control, total**

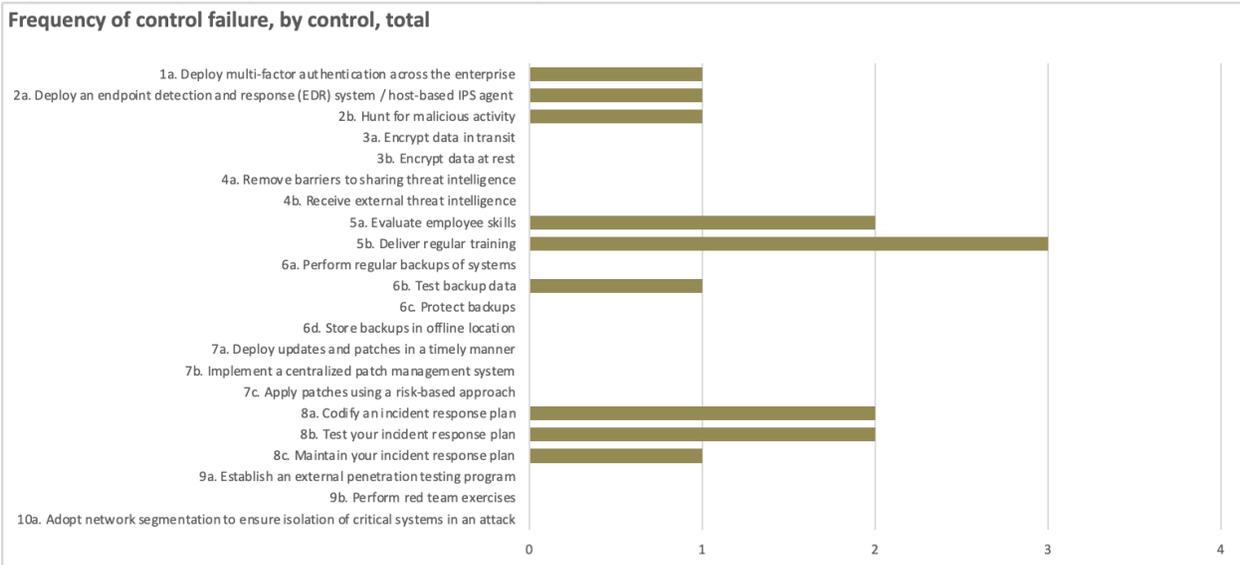

Other controls that failed more than once include "codify an incident response plan" and "test your incident response plan". In many cases, the lack of an incident response plan can extend downtime and lead to higher losses. The controls for MFA, EDR, hunt for malicious activity, test backup data, and maintain and incident response plan all had one attribution that led to a financial loss.

Again, when municipalities have an incident with a loss, they attribute that loss to specific control failures representing that an implemented control failed or that a control was not implemented (up to 5). The financial value of that loss is then allocated evenly across all the implicated controls. Then total losses from each control are summed together to produce a total loss amount per control. This data permits us to see which control failures are leading to the largest losses across the group.

Figure 10 shows the financial losses attributed to each of the sub-controls. The controls related to user training have USD 247,000 of attributed losses followed closely by incident response failures with USD 244,000 in losses. EDR failures are also high and account for USD 112,000 of the municipal losses. Finally, backup failures led to smaller losses of USD 15,000 while MFA was attributed USD 12,000. User training and incident response failures alone accounted for 78% of the total of USD 628,000 in losses across the group.

**Figure 10: Attributed losses from control failures, total, all municipalities, USD**

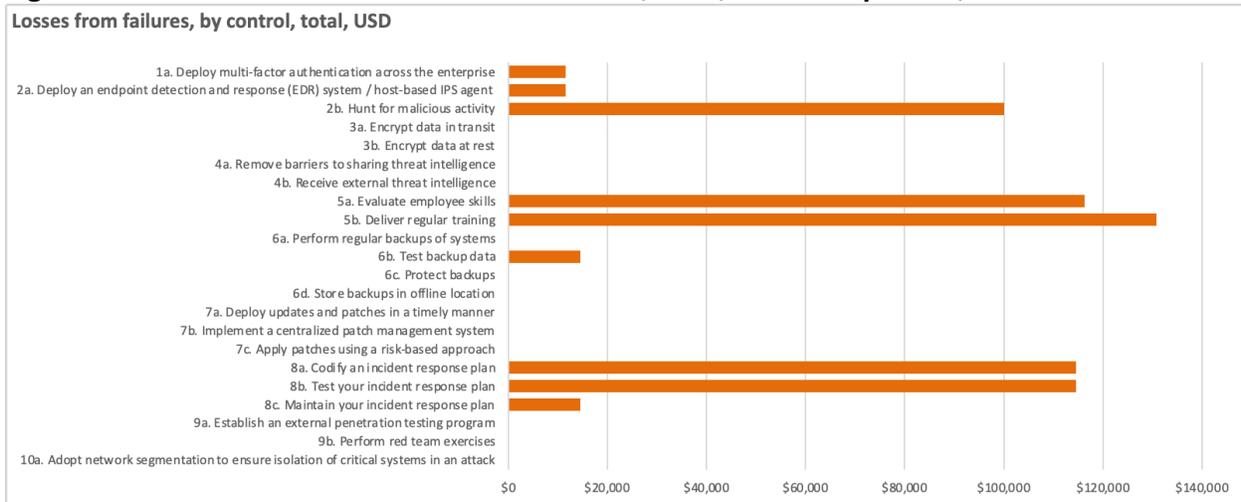

These benchmarking results provide some insight into how municipalities see their own defensive posture, which controls fail the most frequently, and the losses attributed to specific control failures. These results are a snapshot over time for a selection of municipalities in one state. Benchmarks such as these done across states, and particularly over time, can illuminate the evolution of critical infrastructure security, track continuous improvement over time, and highlight areas for specific targets for policymakers.

## Risk modeling results

The benchmarking results can also be used to build models that forecast cyber risk based on an organization's own defense profile. The results in this section bring together data on frequency, the Defense Gap Index scalar, and losses to create the cyber risk model. This section provides the results for each of the subsections and then the results of the overall risk model. Details of the risk modeling approach are available in Annex 2.

### Risk results

The final step is producing the risk forecasts based on the model in equation 1 with the findings from the data collection shown in equation 2. The results of these risk forecasts are shows below in Figures 11 and 12. Each of the figures show how forecasted risk increases or decreases with changes in the Defense Gap Index.

(EQ 1)        $Frequency * DefenseGapIndex * Impact = AnnualRisk$
(EQ 2)        $(0.016) * DefenseGapIndex * (\$157,000) = AnnualRisk$

Figure 11 shows annual expected risk based on a municipality's defense posture. The average risk of USD 2,523 for the average level of protection reflects the "fair price" for an insurance premium based on a pool with all 83 municipalities. In insurance parlance, this fair price is the equivalent of the expected loss for the pool, but does not include other internal costs, external costs, economic profit needs, and capital costs that the insurance provider incurs to run its business. This means that the actual premium would need to be somewhat higher than the calculated expected loss for the insurance company to operate. The "fair price" calculation also assumes that all costs would be covered in the case of an incident, but that is typically not the case as there are exclusions and deductibles that lead to less than full coverage. The "fair price" calculations are imprecise, but they still provide a good guide for organizations to evaluate insurance offers.

**Figure 12: Annual cyber risk forecasts by net weighted security control deviation from group**

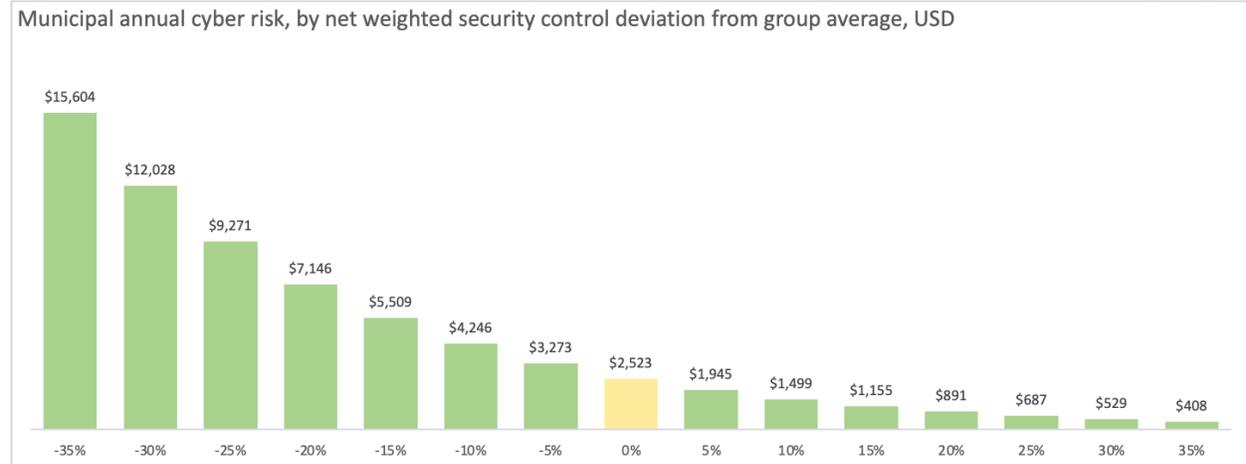

One of the key findings for policymakers is that focusing efforts on organizations with the lowest security levels has a much higher return on investment than similar incremental improvements for organizations with above-average security levels. Our model shows that the risk for organizations with lower security is much higher than the average, with expected losses of over USD 15,000 per year if the weighted net security gap is 35% larger than the average organization. Improving the security of organizations with the weakest defense postures would quickly improve the risk profile of the overall pool given the distribution of losses. It is also worth noting that improving the security of organizations that all already have net weighted security postures that are better than average does have positive returns, but at much lower levels than improving the security of firms that are much lower than average.

Figure 12 shows the same trend but forecasts the financial size of individual security incidents based on the security posture relative to the average. An organization with the average security posture could expect an incident size of USD 157,000 when there is a successful attack. However an organization with a weighted net security gap that puts it 30% below average would expect an incident to cost USD 749,000 – nearly 5 times the average.

**Figure 12: Forecasted incident sizes by net weighted security control deviation from group**

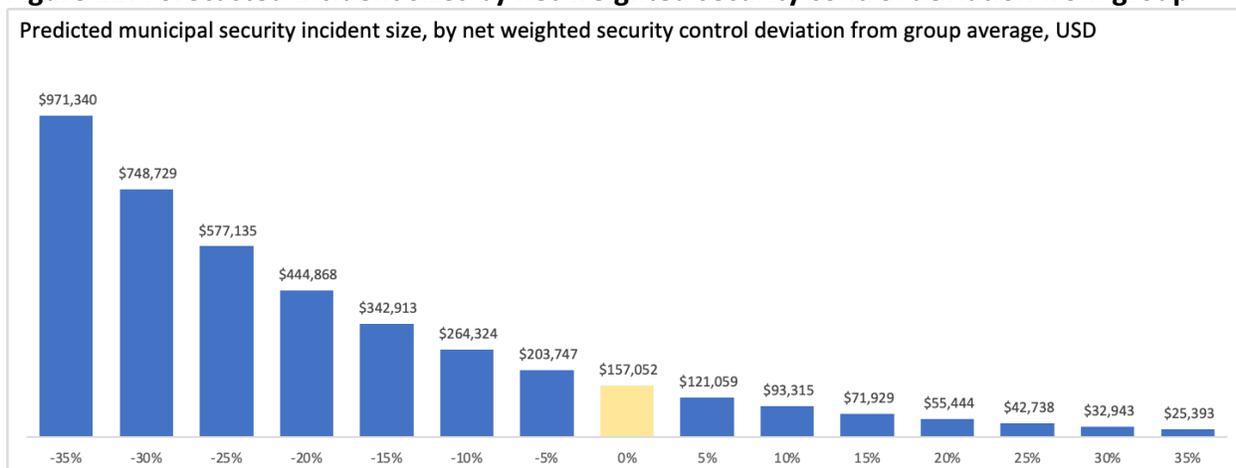

Predicted municipal security incident size, by net weighted security control deviation from group average, USD

*Leveraging the results*

The security and privacy guarantees of the multi-party computation approach we use in SCRAM mean that we as researchers cannot see the input data or the individual security posture of the organizations that submit data into the platform. We only see the aggregated results which can be used to build models such as the ones above.

We do, however, build tools that we distribute back to the participants with these calculated models that allow them to privately evaluate their own risk forecasts based on the private data they submitted. We, as the operators of SCRAM, do not have access to these individual forecasts that the participants can put together using their own private data, but may in the future run a second computation to gather up the risk results and have a better picture off the overall risk landscape for the pool of participants.

## Policy implications

The results of the computation with 83 municipalities provide some important insights at both the state and national level for supporting municipal cybersecurity and investing effectively in security controls and training. This section provides policy considerations and recommendations at both levels emerging from the analysis.

### Securely built risk models provide valuable information to policymakers

One of the key findings of this research is that secure and private data collections can provide valuable information to policymakers who otherwise may lack the information they need to prioritize cyber policy interventions and track their effectiveness. The relatively small amount of data that was collected from each municipality was enough to benchmark defenses, identify gaps, prioritize controls, and build risk models. This data has been too sensitive to disclose in the past, but new private computation techniques have made it possible to produce aggregated results across a peer group without any data disclosure. This makes new data sets available to policymakers that were never available before. With this pilot, we show what is possible with data collected from these new secure computation technologies and hope to convert the results into usable risk metrics.

## The need for a standardized language for talking about and measuring risk

In the past, the reluctance to pool information meant there were no standardized definitions for cyber risk because the data was always considered too sensitive to share. With the development of secure computational techniques, there is now a need to harmonize the language that organizations and policymakers use to talk about cyber risk. For example, there is even confusion about the difference between an "event" and an "incident", with the wording sometimes used interchangeably, and other times used to differentiate between common anomalies versus infrequent larger issues. Now that secure computation is available, we need common set of language and tools for defining and measuring cyber risk.

## Organizations with low security levels benefit most from policy interventions

Another key finding of the paper is that policy interventions targeting organizations with the lowest security posture provide highest return on investment in terms of risk reduction. Policy interventions that focus on bringing up all organizations to a base level of security will have more overall risk reduction than marginally improving the security of municipalities that have above-average security levels.

## Need for regular benchmarking and dynamic improvement

Our focus of this pilot research is on the operators of critical infrastructure that support the quality of life we enjoy in communities. Municipalities lack the same level of resources as large private firms to defend their networks, but these municipal operations are vital to the population and can have life or death consequences if disrupted.

As is noted earlier in the paper, typically data on cyber security defense posture, incident frequencies, and loss data are too sensitive to share. As a result, we have very little insight to just how vulnerable municipalities and the critical infrastructure they operate are to cyber-attacks.

This benchmarking exercise offers us a detailed view into the maturity level of security controls in one US state throughout its municipalities, and the need to have regular benchmarking exercises to identify security gaps, evaluate policy interventions, and allocate resources effectively. The data that we collect as part of this research is only one snapshot in time, but cyber security is an ever-evolving challenge for municipalities. The threat actors modify their approaches, and defenses need to evolve to counter them. As a result, one static snapshot in time is not enough for the dynamic improvements that municipalities must make to defend their assets. As a result, we believe that this research highlights the need to have benchmarking exercises on a regular basis to evaluate the progression of defenses and understand the risk that municipalities are facing. Having a cross-section and time series of data allows a much richer analysis and better policy recommendations.

## Working toward a definition of reasonable cybersecurity for municipal governments

Security benchmarks for municipalities paired with information on actual losses helps policymakers understand and work toward a definition of reasonable cybersecurity for municipal governments. Municipalities would benefit from advice on which controls should be implemented and in which prioritized order.

## Dynamic Improvement and insurance-based schemes

The decentralized nature of municipal government and decision-making in the United States makes it difficult to coordinate effective response to risks such as cybersecurity threats. There is a precedent for aligning the incentives of local governments to reducing systemic risk through an insurance-based scheme: the National Flood Insurance Program (FEMA, 2020), in particular the Community Rating System. The National Flood Insurance Program, administered by the Federal Emergency Management Agency (FEMA), offers federally backed flood insurance for homeowners and businesses located in high and moderate risk flooding areas. The NFIP was created in 1968 by the passage of the National Flood Insurance Act [3]. The enactment of the National Flood Insurance Act was itself a response to a lack of market-based flood insurance for high-risk areas and mounting federal flooding disaster assistance costs (NAIC, 2022). As part of remaining eligible to participate in the National Flood Insurance Program, participating communities, as they are known, must fulfill minimum NFIP requirements regarding proper floodplain management practices.

> "The Community Rating System was created to encourage communities to establish sound programs that recognize and encourage floodplain management activities that exceed the minimum NFIP requirements. By conducting mitigation and outreach activities that increase safety and resilience, including CRS credits for regulating to higher standards, communities can earn credits and discounts (up to 45 percent within the Special Flood Hazard Area) on flood insurance premiums for property owners." (FEMA, 2023a)

Participating communities that meet the higher CRS standards are eligible for reduced flood insurance premiums. The CRS Coordinator Manual (FEMA, 2023b) details a series of flood preparedness goals, and relates fulfillment of those goals with a reduction of flood insurance premium prices, up to a 45% discount. Each community participating in CRS must attest yearly that it is implementing its stated flood mitigation policies in a process known as recertification. Communities are subject to a lengthier verification process at the conclusion of three-year or five-year cycles, depending on community classification and risk profile.

The Massachusetts Municipal Association, in partnership with the Massachusetts Interlocal Insurance Association, runs a cybersecurity liability insurance program (MIIA, 2023). A potential incentive mechanism for encouraging municipalities to participate in regular benchmarking and improvement in their cybersecurity posture would be to tie favorable insurance rates to continuous improvement on the self-reported SCRAM assessment metrics, combined with regular auditing by the insurance provider to ensure compliance.

## Policy related to specific security controls

This analysis has identified three core areas of weakness among municipal cyber defenses that are responsible for significant losses.

The largest impact area is employee training and evaluating employee skills. Failures of these two controls accounted for 39% of all losses seen across the group, and three out of four incidents blamed employee training and evaluating skills as core reasons for the ultimate financial loss. These results are common across other computations we have run in other sectors.

There is a nearly uniform distribution of responses across the questions related to evaluating employee skills and delivering regular training (Figures 13 and 14). The good news is that 36% of municipalities say they have fully implemented regular cyber training, and 24% reported a fully implemented program for evaluating employee skills. More concerning is the bottom end of the responses where 16 municipalities do not deliver regular training and 21 municipalities do not evaluate employee skills.

**Figure 13: Distribution of responses for evaluating employee skills, counts**

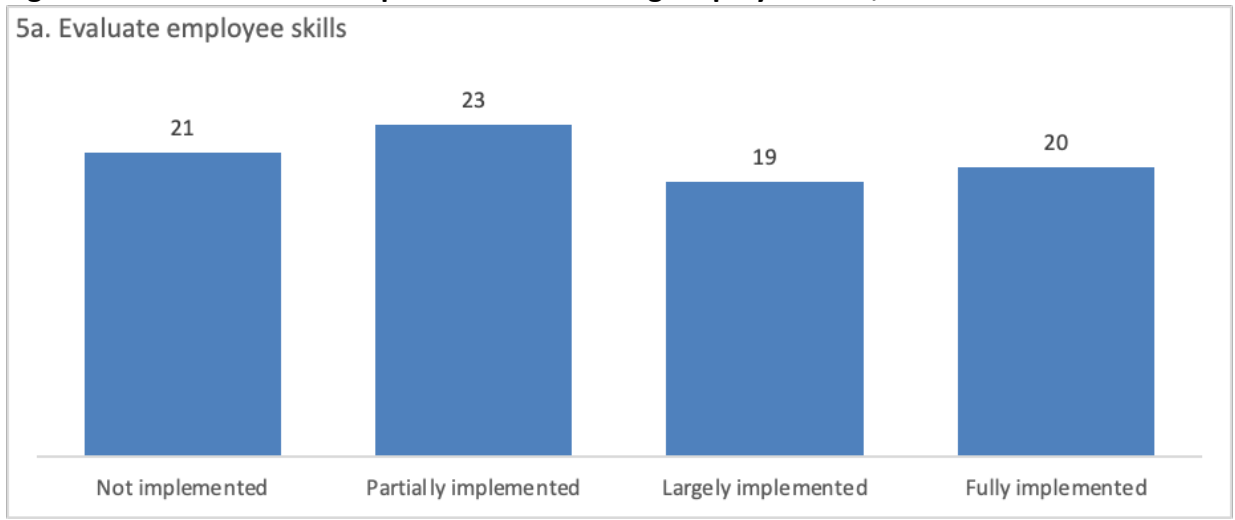

**Figure 14: Distribution of responses for delivering regular training, counts**

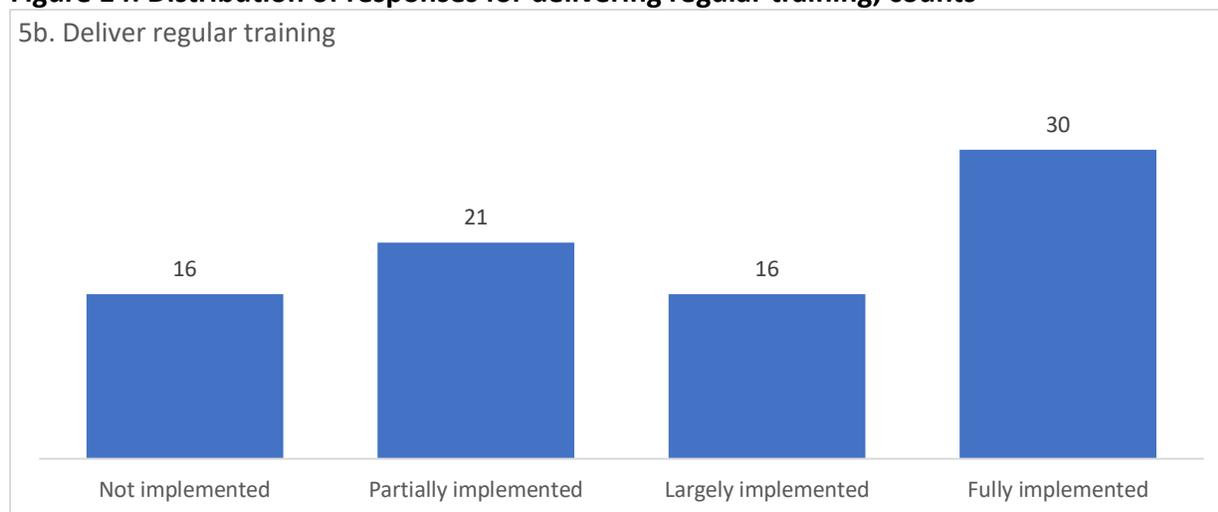

Employee skills is a critical area of weakness that all sectors need to address because human errors remain a core entry method for threat agents. In talking with municipalities, we realized that many organizations did not have the funding to run extensive training and felt they did not have the adequate skills to develop materials. This points to a potential role for state and national governments to help bring together resources that can be leveraged for employee training to defend these municipalities.

Incident response also had large losses associated with it that represented 39% of all reported losses across the municipalities. The largest losses were associated with codifying and incident response plan and testing that plan once it is ready. What was surprising is that the state has detailed resources for incident response planning that are available to the municipalities. Additionally, the insurance provider also has extensive incident response resources that are available at any time to the municipalities. This may imply that resources are available, but there is a lack of awareness about them.

Finally, there were significant losses associated with detecting malicious activity, either through hunting for malicious activity, or by deploying an endpoint detection and response system. These detection failures accounted for 18% of all reported losses across the group. One of the reasons that EDR and hunting for malicious activity or critical is that the costs associated with an incident increase the longer it takes to detect. Systems that can quickly detect anomalous activity and allow organizations to remedy the problem will reduce the impact of events.

### Awareness raising

As mentioned above, municipalities often were unaware of the resources that were available to them at the state level, and they did not know about support they could receive from the insurance provider when faced with an incident. One recommendation coming out of this research could be for awareness raising among municipalities about the resources currently available at their disposal.

## Materials / training at the state level

Employee training and evaluating employee skills, as mentioned before, is a common failure point across the municipalities but also in other sectors of the economy. There may be a role for states to develop resources and training for cyber hygiene that can be provided to employees. It may make more sense to have high quality materials available for everyone than for each organization to try and develop their own. Municipalities could take elements developed at the state level and tailor them specifically to their own needs. Among the municipalities, there are many with similar roles and similar critical infrastructure portfolios, so even subgroups of municipalities could work together on unified training materials with state support.

## Evaluating the results of existing programs

Over the last decade, various programs have been implemented by states to support municipalities in building resiliencies. Of note among them include the Michigan Cyber Civilian Corps (MiC3) (NGA, 2017) , "a group of trained, civilian technical experts who individually volunteer to provide rapid response assistance to the State of Michigan in the event of a critical cyber incident". The Corps is composed of IT professionals who already hold certain IT and security certifications and participate in additional training regarding the nuances of working in the public sector. Other programs include regional CISO programs, where an individual paid by the state splits CISO responsibilities for multiple municipalities at once (StateTech, 2020). A common resource is the provision of state-provided employee cybersecurity training such as those run in Massachusetts (Commonwealth of MA, 2023). Crucially, data about the efficacy about these well-intentioned state programs has not been routinely collected or made available making It challenging to establish which programs are the best allocation of scarce state resources in service of municipal cybersecurity.

Without the ability to benchmark progress, there is no way to gauge the impact of these programs and evaluate their effectiveness. By building an ongoing benchmarking program, states could evaluate the efficiency and effectiveness of these programs in different groups of municipalities. Ultimately, this would help states prioritize their security investments toward interventions with the most impact.

While our focus in this paper has been on municipal cyber security, there are also lessons that we can learn from this data at the national or broader level. For example, there are no standardized risk measures or benchmarks of control, maturity across municipalities, states, or even countries. One policy consideration could be encouraging standardized risk measures and security benchmarking across states, and potentially across nations that could be used for apples-to-apples comparisons. The larger the group of comparable data, the quicker that lessons learned can be disseminated and implemented across the organizations.

There is also a need that goes beyond the state level for governments to encourage vendors and government agencies to work together on addressing the largest causes of cyber losses. For example, how can the broader public and private community work together to improve cyber education for employees and reduce the number of significant incidents? Governments are large buyers of cyber security tools and they may have some leverage with suppliers to help create tools, standards, and use cases that work across different sectors.

Finally, benchmarking and standardized measures could be promoted at not just the state and national level, but also with relevant international organizations to provide more opportunities for broader benchmarking, risk aggregation assessment, and research into the overall impact of cyber security on the economy. Organizations such as the OECD could play an important role.

## Next steps and future research

Future research on municipal cyber risk could track changes over time and how those changes affect loss outcomes and overall risk measures. The research could also be expanded to include and compare municipalities from other states.

Additional research is needed to understand the relationship between defensive posture and losses to do accurate risk modeling. Future research could develop new ways to test for these correlations in a private and secure manner.

The frequency of attacks and successful attempts are likely dependent on elements such as the municipality's size or its attractiveness as a target. Our analysis in this paper takes the frequency of attacks as static, but future research could examine how to model the frequencies more precisely.

## Bibliography

Baltimore Sun, 2019a. Baltimore estimates cost of ransomware attack at $18.2 million as government begins to restore email accounts [WWW Document]. Baltimore Sun. URL https://www.baltimoresun.com/maryland/baltimore-city/bs-md-ci-ransomware-email-20190529-story.html (accessed 9.10.23).

Baltimore Sun, 2019b. Home sales are held up; Baltimore ransomware attack cripples systems vital to real estate deals [WWW Document]. Baltimore Sun. URL https://www.baltimoresun.com/maryland/baltimore-city/bs-md-ci-ransomware-home-sales-20190514-story.html (accessed 9.10.23).

Boustead, A., Shackelford, S., 2020. State of Hoosier Cybersecurity 2020.

Commonwealth of MA, 2023. Municipal Cybersecurity Awareness Grant Program | Mass.gov [WWW Document]. Municipal Cybersecurity Awareness Grant Program. URL https://www.mass.gov/municipal-cybersecurity-awareness-grant-program (accessed 9.10.23).

Courtney, R.H., 1977. Security risk assessment in electronic data processing systems, in: Proceedings of the June 13-16, 1977, National Computer Conference on - AFIPS '77. Presented at the the June 13-16, 1977, national computer conference, ACM Press, Dallas, Texas, p. 97. https://doi.org/10.1145/1499402.1499424

de Castro, L., Lo, A.W., Reynolds, T., Susan, F., Vaikuntanathan, V., Weitzner, D., Zhang, N., 2020. SCRAM: A Platform for Securely Measuring Cyber Risk. Harvard Data Science Review 2. https://doi.org/10.1162/99608f92.b4bb506a

Deloitte, NASCIO, 2018. 2018 Deloitte-NASCIO Cybersecurity Study - States at Risk: Bold Plays for Change. NASCIO. URL https://www.nascio.org/resource-center/resources/2018-deloitte-nascio-cybersecurity-study-states-at-risk-bold-plays-for-change/ (accessed 9.10.23).

Deloitte, NASICO, 2020. 2020 Deloitte-NASCIO Cybersecurity Study - States at Risk: The Cybersecurity Imperative in Uncertain Times. NASCIO. URL https://www.nascio.org/resource-center/resources/2020-deloitte-nascio-cybersecurity-study-states-at-risk-the-cybersecurity-imperative-in-uncertain-times-2/ (accessed 9.10.23).

Eling, M., Wirfs, J., 2019. What are the actual costs of cyber risk events? European Journal of Operational Research 272, 1109–1119. https://doi.org/10.1016/j.ejor.2018.07.021

FEMA, 2023a. Local Government Officials - Floodplain Management Resources | FEMA.gov [WWW Document]. URL https://www.fema.gov/floodplain-management/manage-risk/local (accessed 9.10.23).

FEMA, 2023b. Community Rating System Coordinator's Manual [WWW Document]. URL https://www.fema.gov/sites/default/files/documents/fema_community-rating-system_coordinators-manual_2017.pdf (accessed 9.10.23).

FEMA, 2020. Laws and Regulations | FEMA.gov [WWW Document]. URL https://www.fema.gov/flood-insurance/rules-legislation/laws (accessed 9.10.23).

Freund, J., Jones, J., 2014. Measuring and Managing Information Risk: A FAIR Approach. Butterworth-Heinemann.

MIIA, 2023. MIIA - About MIIA [WWW Document]. URL https://www.emiia.org/about (accessed 9.10.23).

Mukhopadhyay, A., Chatterjee, S., Bagchi, K.K., Kirs, P.J., Shukla, G.K., 2019. Cyber Risk
Assessment and Mitigation (CRAM) Framework Using Logit and Probit Models for Cyber
Insurance. Inf Syst Front 21, 997–1018. https://doi.org/10.1007/s10796-017-9808-5

NAIC, 2022. Flood Insurance/National Flood Insurance Program (NFIP) [WWW Document]. URL
https://content.naic.org/cipr-topics/flood-insurancenational-flood-insurance-program-
nfip (accessed 9.10.23).

NGA, 2017. National Governors Association Memo: Building a Civilian Cyber Corps (Memo).

Norris, D.F., 2021. A Look at Local Government Cybersecurity in 2020 [WWW Document]. A
Look at Local Government Cybersecurity in 2020. URL https://icma.org/articles/pm-
magazine/look-local-government-cybersecurity-2020 (accessed 9.10.23).

Norris, D.F., Mateczun, L., Joshi, A., Finin, T., 2021. Managing cybersecurity at the grassroots:
Evidence from the first nationwide survey of local government cybersecurity. Journal of
Urban Affairs 43, 1173–1195. https://doi.org/10.1080/07352166.2020.1727295

Norris, D.F., Mateczun, L., Joshi, A., Finin, T., 2019. Cyberattacks at the Grass Roots: American
Local Governments and the Need for High Levels of Cybersecurity. Public Administration
Review 79, 895–904. https://doi.org/10.1111/puar.13028

Preis, B., Susskind, L., 2022. Municipal Cybersecurity: More Work Needs to be Done. Urban
Affairs Review 58, 614–629. https://doi.org/10.1177/1078087420973760

Reuters, 2018. Atlanta officials reveal worsening effects of cyber attack. Reuters.

Spiewak, R., 2022. Overlooking the Little Guy: An Analysis of Cyber Incidents and Individual
Harms (Thesis). Massachusetts Institute of Technology.

Spiewak, R.L., Reynolds, T.W., Weitzner, D.J., 2021. Ransomware Readiness Index: A Proposal to
Measure Current Preparedness and Progress Over Time (Working Paper).

State of Louisiana, 2019. Cybersecurity Incident Resources | Office of Governor Jeff Landry
[WWW Document]. URL https://gov.louisiana.gov/page/cybersecurity-incident-
resources- (accessed 1.31.24).

StateTech, 2020. Michigan's CISO as a Service Boosts Local Cybersecurity [WWW Document].
Technology Solutions That Drive Government. URL
https://statetechmagazine.com/article/2020/04/michigans-ciso-service-boosts-local-
cybersecurity (accessed 9.10.23).

US Census, 2023. 2022 Census of Governments [WWW Document]. Census.gov. URL
https://www.census.gov/data/tables/2022/econ/gus/2022-governments.html (accessed
1.31.24).

US CISA, 2023. Report Ransomware | CISA [WWW Document]. URL
https://www.cisa.gov/stopransomware/report-ransomware (accessed 1.31.24).

US DHS, 2023. DHS Announces Additional $374.9 Million in Funding to Boost State, Local
Cybersecurity | Homeland Security [WWW Document]. URL
https://www.dhs.gov/news/2023/08/07/dhs-announces-additional-3749-million-
funding-boost-state-local-cybersecurity (accessed 1.31.24).

US White House, 2023. Biden-Harris Administration Launches New Efforts to Strengthen
America's K-12 Schools' Cybersecurity [WWW Document]. The White House. URL
https://www.whitehouse.gov/briefing-room/statements-releases/2023/08/07/biden-
harris-administration-launches-new-efforts-to-strengthen-americas-k-12-schools-
cybersecurity/ (accessed 1.31.24).

US White House, 2021a. Executive Order on Improving the Nation's Cybersecurity | The White House 24.

US White House, 2021b. White House Memo to Corporate Executive and Business Leaders on Ransomware.

US White House, 2018. The Cost of Malicious Cyber Activity to the U.S. Economy.

Questions answered as one of four levels of implantation: "Not", "Partially", "Largely", "Fully"
1a. Deploy multi-factor authentication across the enterprise
2a. Deploy an endpoint detection and response (EDR) system / host-based IPS agent
2b. Hunt for malicious activity
3a. Encrypt data in transit
3b. Encrypt data at rest
4a. Remove barriers to sharing threat intelligence
4b. Receive external threat intelligence
5a. Evaluate employee skills
5b. Deliver regular training
6a. Perform regular backups of systems
6b. Test backup data
6c. Protect backups
6d. Store backups in offline location
7a. Deploy updates and patches in a timely manner
7b. Implement a centralized patch management system
7c. Apply patches using a risk-based approach
8a. Codify an incident response plan
8b. Test your incident response plan
8c. Maintain your incident response plan
9a. Establish an external penetration testing program
9b. Perform red team exercises
10a. Adopt network segmentation to ensure isolation of critical systems in an attack

Questions answered as a number:
11a. Number of significant incidents over three years (sum of 2019, 2020, 2021) (see note*)
11b. Total cyber losses for all incidents combined over 3 years, US$, (sum of 2019, 2020, 2021)

Questions answered via a selection:
12: If incidents are listed, the respondent must select up to 5 control failures that led to the loss.

A screenshot of the questionnaire in an Excel spread sheet is shown below in Figure 15.

# Figure 15: Screenshot of the MIT RRI Questionnaire for municipalities

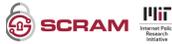

| Not ready | |
|---|---|
| 167 | Checksum |

**Step 1: Population estimate**

| | | Input number | Status |
|---|---|---|---|
| Population | 0a. The population of your municipality (rough estimates okay) | | Need data |

**Step 2: Maturity levels**

| Category | Control | Mark with an "x" | | | | Status |
|---|---|---|---|---|---|---|
| | | Not Implemented | Partially Implemented | Largely Implemented | Fully Implemented | |
| 1. MFA | 1a. Deploy multi-factor authentication across the enterprise | | | | | Need Data |
| 2. EDR | 2a. Deploy an endpoint detection and response (EDR) system / host-based IPS agent | | | | | Need Data |
| | 2b. Hunt for malicious activity | | | | | Need Data |
| 3. Encryption | 3a. Encrypt data in transit | | | | | Need Data |
| | 3b. Encrypt data at rest | | | | | Need Data |
| 4. Empowerment | 4a. Remove barriers to sharing threat intelligence | | | | | Need Data |
| | 4b. Receive external threat intelligence | | | | | Need Data |
| 5. Training | 5a. Evaluate employee skills | | | | | Need Data |
| | 5b. Deliver regular training | | | | | Need Data |
| 6. Backup | 6a. Perform regular backups of systems | | | | | Need Data |
| | 6b. Test backup data | | | | | Need Data |
| | 6c. Protect backups | | | | | Need Data |
| | 6d. Store backups in offline location | | | | | Need Data |
| 7. Patch | 7a. Deploy updates and patches in a timely manner | | | | | Need Data |
| | 7b. Implement a centralized patch management system | | | | | Need Data |
| | 7c. Apply patches using a risk-based approach | | | | | Need Data |
| 8. Incident response | 8a. Codify an incident response plan | | | | | Need Data |
| | 8b. Test your incident response plan | | | | | Need Data |
| | 8c. Maintain your incident response plan | | | | | Need Data |
| 9. Check the work | 9a. Establish an external penetration testing program | | | | | Need Data |
| | 9b. Perform red team exercises | | | | | Need Data |
| 10. Segment | 10a. Adopt network segmentation to ensure isolation of critical systems in an attack | | | | | Need Data |

**Step 4: Identifying failures**

| Mark with an "x" | | |
|---|---|---|
| Select up to 5 controls that failed during incidents that incurred the greatest financial losses. This includes either a control failure, or a lack of a control that could have prevented the loss. | | Status |
| 1a | | Not needed |
| 2a | | Not needed |
| 2b | | Not needed |
| 3a | | Not needed |
| 3b | | Not needed |
| 4a | | Not needed |
| 4b | | Not needed |
| 5a | | Not needed |
| 5b | | Not needed |
| 6a | | Not needed |
| 6b | | Not needed |
| 6c | | Not needed |
| 6d | | Not needed |
| 7a | | Not needed |
| 7b | | Not needed |
| 7c | | Not needed |
| 8a | | Not needed |
| 8b | | Not needed |
| 8c | | Not needed |
| 9a | | Not needed |
| 9b | | Not needed |
| 10a | | Not needed |

**Note:** If there were no incidents in 2019, 2020, of 2021 with losses over $1,000, then stop here. Otherwise, continue to Step 3 (below) and Step 4 (to the right)

or ✗

**Step 3: Incidents and losses**

| | | Input number | Status |
|---|---|---|---|
| Incidents | 11a. Number of significant incidents over three years (sum of 2019, 2020, 2021) (see note*): | | None reported |
| Cyber loss estimate (all incidents) | 11b. Total cyber losses for all incidents combined over 3 years, US$, (sum of 2019, 2020, 2021): | | Not needed |

*Definition of a significant incident: Incidents against your organization within the past 3 years with a minimum loss amount of $1,000. Multiple endpoints, network segments, platforms or accounts hit in an attack that require remediation count as 1 incident.

## Data

This section describes how use our secure data collection platform SCRAM to gather data from 83 municipalities in Massachusetts and use them to produce security benchmarks, risk models, and forecasts to understand the state of defenses, evaluate the risk, and provide a new feedback mechanism to evaluate state and national programs.

The data were collected in June 2022 from the municipalities in Massachusetts via a secure questionnaire prepared for the SCRAM platform. A description of each of the data elements is provided below. It is important to note, however, that the results from a multi-party computation are only aggregate figures and cannot be linked back to an individual organization that submitted data. The increase in privacy and security comes at the expense of no longer being able to see the individual inputs into a data aggregation.

All data are self-reported by municipalities, and participants are trained during a session on how to fill out the forms and estimate the data based on their own experiences. This training step is necessary to improve the comparability of the responses across firms.

The collected data that are used in this analysis include:

- **Maturity level:** 22 questions on the maturity level of controls (see Annex 1)
- **Incident count:** 1 question on the frequency of incidents
- **Control failures:** Count of the times individual controls failed leading to incidents
- **Financial costs (total):** 1 question on the total cost of incidents
- **Financial costs (control failures):** Data on the attributed costs of incident failures to specific controls

### Data: Maturity levels

The data for this analysis include a set of 22 questions across 10 categories on the maturity of cybersecurity controls related to ransomware. We use the MIT Ransomware Readiness Index (RRI) questionnaire (Spiewak et al., 2021) as the set of controls municipalities use to provide a self evaluation for the computation. The RRI is a distillation of the controls identified by the 2021 White House Executive Order on Cybersecurity (US White House, 2021a) and the US White House memorandum on ransomware (US White House, 2021b) as essential in assessing organizations' cybersecurity readiness. In the RRI, participants are asked to report their status (Not Implemented, Partially Implemented, Mostly Implemented, and Fully Implemented) on 22 controls in 10 categories: Multifactor Authentication, Endpoint Detection and Response, Encryption, Empowerment, Training, Backup, Patching, Incident response, Checking the work, and Segmenting. The answers to these questions are mapped to a 100% scale shown in Table 1.

**Table 1: Implementation levels mapped to percentages for quantification**

| Not implemented | Partially implemented | Largely implemented | Fully implemented |
|---|---|---|---|
| 0% | 33% | 67% | 100% |

Data: Incident counts

The data on incident counts asks organizations for the total number of incidents with a financial loss greater than USD 1,000. The incident count data is the sum across all organizations in each of the five groups, in our case grouped by population size. There were 4 incidents reported across 83 municipalities over the period of 3 years from 2019 to 2021).

Data: Control failures

Organizations must identify between 1 and 5 control failures that are responsible for the losses they report as part of an incident. The respondents are instructed to identity the top 5 controls responsible for all their incidents in cases where they have more than one incident to report. If an organization reports an incident, then they must identify at least one control failure. There are 14 control failures identified across the 4 reported incidents, of which 9 are unique. The list of controls with failures is provider later in Table 3.

Financial costs: Total

The data include information on the total financial losses of all the incidents combined. The total financial losses from the 4 cyber incidents over the 3-year period were USD 628,000.

Financial costs: Controls

The data collection also allocates reported losses across controls that were reported to fail. The total loss for each control is the portion of losses for all incidents that were allocated to failures of that control. Each control has a total loss amount across all the municipalities. There are 9 controls with attributed losses contributed by the 83 municipalities over the 3 years.

Model

For this paper, we propose a risk model that includes a frequency of incidents measure, a Defense Gap Index, and impacts (losses) as the three top-level components determining cyber risk. Assembling these three components together in a model lets us make changes to any of the three and then see how they impact an organization's risk. From a policy perspective, we care about how certain "levers" that affect the each of the three affect the overall quantified risk of the organization.

Our Defense Gap Index is comparable to $Sec_T$ in the CRAM model (Mukhopadhyay et al., 2019), but we have named it differently to capture the dynamic that higher scores of the variable relate to higher risk. As a result, the name Defense Gap Index more accurately captures the dynamics of control variations on risk measures. Bringing the Defense Gap Index (a measure of security posture) up to the top level alongside frequency and impact allows us to work more directly with changes in control strength and their impact on risk outcomes.

In our model, the annual cyber risk for a firm is determined by the historical frequency of attacks in the sector multiplied by a defense gap index scalar and by the average impact of incidents for the sector (Equation 1). Our key modeling assumptions for the municipal risk analysis are detailed in the section below.

(Eq 1) $$Frequency * DefenseGapIndex * Impact = AnnualRisk$$

## Frequency

In our general model, outside factors can influence the frequency of attacks including the sophistication of threat actors, the industry profile, firm size, and the appeal of the protected assets to those wishing to exploit systems.

In this municipal implementation however, we only have access to historical frequency rates based on our limited data collection. We assume that any outside factors (sophistication of actors, etc.) that can influence frequency are constant across the population in the data collection. The frequency data is determined by a SCRAM computation that collects the total number of incidents and their magnitudes across the municipalities over a three-year period and then converts it into an annual frequency rate per municipality. The data are self-reported and private but are checked against information we received from the insurance provider.

Frequency calculation method:
- Count of significant incidents across group over 3-year period / number of firms x 3

Frequency Assumptions:
- All outside factors that can affect frequency such as actor sophistication and assets being protected are assumed to be constant (similar) across all municipalities in the sample.
- The impact of security controls is not captured in frequency. They are picked up in the defense gap index.

## Defense Gap Index

The defense gap index is a way to adjust risk levels based on how an organization's control maturity compares to its peer group. The index is a scaler that captures deviations of weighted security control maturity levels from group averages and applies these deviations to a nonlinear model of losses to produce a scalar that impacts the risk forecasts. The innovation of the Defense Gap Index is that it uses actual loss data and attributions to specific control failures from the group to weight each of the controls in the index. Then, variations away from the average are scaled according to these derived weights.

Producing the Defense Gap Index is a multi-step process as follows:

**Step 1: Allocate overall category weights between controls groups with and without losses**

In this first step we decide how much importance to place on control failures that led to losses, the "loss group", compared to controls that were not implicated in loss events, the "no-loss group".

- In the municipality context, we allocate 85% of the total weight to the loss group which has 9 controls. We assign the remaining 15% to the no-loss group that has 13 controls. This breakdown of 85/15 was done to ensure that all controls with actual losses have higher weights assigned than controls without losses.

**Step 2: Allocate individual control weights within loss and no-loss groups**

In the second step, we decide how the weights should be subdivided across controls in the loss and no-loss groups. We prorate the 85% total weight in the loss group by the relative losses of individual controls. We assign equal weights across the 15% total weight in the no-loss group (See Figure 16 and Table 3).

- For the municipalities, controls in the loss group receive a prorated share of the overall 85% weight for the indicator based on their total losses across the group. The controls in the no-loss group are all weighted equally across the remaining 15% of total weight.

**Figure 16: How weights are applied to controls with and without attributed losses**

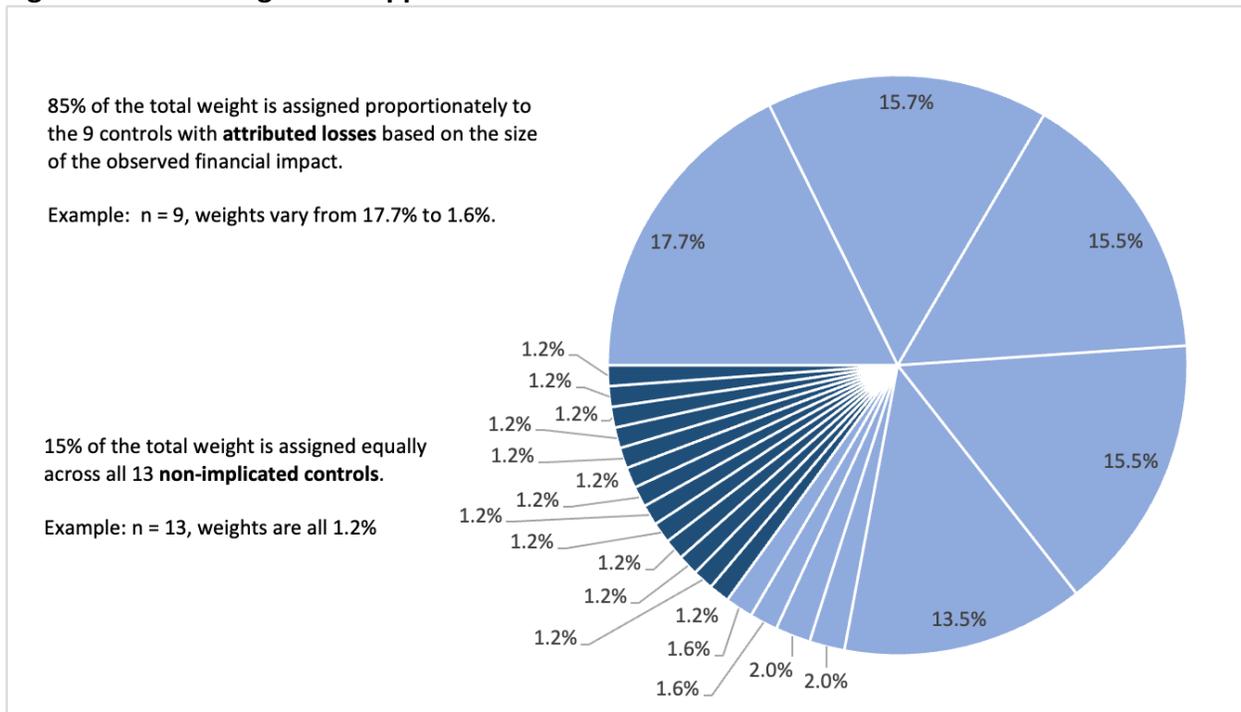

Note: The specific control weights and names are provided in Table 3.

**Step 3: Calculate the net weighted deviation from the group average**

The third step looks for maturity differences from the group average and then weights them based on the control weights in Step 2. The differences are expressed as percentage differences from the group average, then then these are multiplied by the individual control weights and then summed to create a net deviation score.

(Eq 3)
$$NetWeightedDeviation = \sum_{n=1}^{i}((OwnScore_n - GroupAverage_n) * ControlWeight_n)$$

**Step 4: Create a loss model fitting observed loss data to the net weighted deviation for controls**

The fourth step evaluates and models the distribution of observed losses over the net weighted deviation for controls from Step 3. We assume that higher security (positive net weighted variations) corresponds to lower losses. We map the largest losses to the lowest security levels (e.g. 30% below average) and the smallest losses to the higher security levels (e.g. 20-30% above average). We also have the average loss and the average security level. With these 3 or more points in place, next we fit a function to the observed data that will be used to create the Defense Gap Index scalar. The loss data typically follow an exponential function.

- With this municipal data, we derived the following function to map actual loss data do security variances. A detailed derivation of this value is provided in results section.

(Eq 4) $$y = e^{-5.206x}$$

**Step 5: Calculate the Defense Gap Index scaler:**

When we collect data on incident frequencies and impact, the resulting numbers correspond to the average security posture across the sampled group. The Defense Gap Index converts deviations from this average into a scalar that changes the overall risk outcome.

The final step produces the Defense Gap Index scaler that is greater than 1 when maturity levels are lower than average, and less than 1 when an organization's defenses are better than their peers. The size of the scaler is dependent on the weights assigned in earlier stages, so gaps or areas of strength in one do not necessarily compensate 1:1 for gaps or areas of strength in other controls of a similar magnitude.

- The equation for the defense gap index scaler for municipalities

(Eq 5) $$DefenseGapIndex = e^{-5.206*NetWeightedDeviation}$$

The average maturity level of controls across the group is normalized at 1 as the base case against which an organization's own maturity level is evaluated. Organizations with better security than the average receive a lower gap score, while those with worse security receive a higher gap score.

Defense Gap Index Calculation:
- Step 1: Allocate overall category weights between controls groups with and without losses
- Step 2: Allocate individual control weights within loss and no-loss groups
- Step 3: Calculate the net weighted deviation from the group average
- Step 4: Create a loss model fitting observed loss data to the net weighted deviation for controls
- Step 5: Calculate the Defense Gap Index scaler

Defense Gap Index Assumptions

- We assume that any deviations away from the maturity average for controls with reported losses should be given more weight than controls that had no reported losses across the data collection.
- The assigned 85% and 15% weighting reflect our view that controls with observed losses should have the highest weighting, but that controls without losses also play a role and need to be included. This is because successful controls may not have losses precisely because they are stopping the progression of attacks.
- When creating the loss-fitting model, we assume that the largest losses are associated with the lowest security levels. This may not actually be the case, but we believe it is a safe assumption for these modeling purposes. Future data collections will allow us to test for correlations between security levels and losses.

## Losses

Loss data are pulled from the aggregated data in the SCRAM computation. SCRAM outputs the sum of all losses and the average loss per incident. The research team cannot see individual entries, but we do create flags for different ranges of answers to understand the distributions of answers. For example, we could count the number of losses between USD 5,000 and USD 50,000 in one category, and between USD 50,000 and USD 500,000 in another. This provides us with visibility into the distribution of losses that is used to model the Defense Gap Index.

Loss calculations
- Sum of the losses from all reported incidents
- The average loss per incident from the total losses divided by the count of incidents
- A distribution of losses based on broad loss ranges

Loss assumptions
- We assume that the data submitted by firms on their losses is in the general range of actual amounts, but they do not need to be exact. We provide a spreadsheet to all participants that helps them estimate losses for different kinds of incidents.

## Results

### Frequency results

The municipal data put into the SCRAM data collection cover three years (2019, 2020, 2021) for 83 municipalities. Over the course of the three-year period, there are four incidents, which produces an expected frequency a single municipality having an expected successful cyberattack with losses greater than USD 1,000 once every 62.5 years, or a 0.016 chance in any given year. The frequency value should be interpreted with caution over long-term forecasts because tactics and techniques constantly evolve, but it does offer the best frequency estimate we have since it is based on the most recent loss events.

### Defense gap index results

The Defense Gap Index is simply a scaler based on the security state of the municipality that will increase or decrease the risk forecast based on deviations from the group average. We model a band of +/- 30% from the average defense posture. As mentioned earlier, we assume that lower security levels result in higher losses when there is an incident.

We use the losses from the benchmarking exercise to estimate the top and bottom ends of our loss range, with some additional headroom on the top end given our low number of incidents in the sample. The benchmarking data from the SCRAM computation gives us several points we can use to formulate a trendline. The data has one loss of USD 500,000 more, and three losses under USD 100,000. One of the smaller losses is greater than USD 50,000, and the remaining two are below that amount. We also know the average loss (USD 157,000) corresponds to the average security maturity across the group. We assume that organizations with security maturity levels that are 30% higher than the average have only a small risk of losses, while organizations with security levels that are 30% below average are more likely to have the largest losses observed, in our case greater than USD 500,000 per incident.

We use a weighted measure for the Defense Gap Index that assigns higher weights to controls exhibiting large losses and lower weights to controls with small or no losses across the group. We assign 85% of the total weight to security gaps for controls that have losses. We prorate the weights on each of the variable in the "loss group" according to the magnitude of their observed losses. The ratio of 85/15 aligns best to the data so that the lowest weights on controls with losses are higher than the weights assigned to each of the controls in the no-loss group. For example, the control "2a. Deploy EDR" has USD 11,529 in reported losses across the group and receives a weight of 1.6%, while "3a. Encrypt in transit" has no losses but still receives 1.2% of the total weight.

Table 3 shows the observed losses and the prorated weights assigned that are used to calculate the Defense Gap Index scalar. One column shows what the weighting would be if all controls were weighted equally, while the next column shows the prorated control rates weighted by reported losses.

It is worth noting that the prorated control weights are very high for the controls that had large reported losses. Training and evaluating employee skills together account for over 34% of the total weight applied to defensive posture. Organizations with poor training or skill evaluations will have much higher Defense Gap Index numbers and corresponding high-risk forecasts.

**Table 3: Observed losses and prorated control weights for the defense gap index, USD**

| Control | Observed losses | Equal control weights | Prorated control weights by losses: 85% prorated across losses 15% equally across non-losses |
|---|---|---|---|
| 5b. Deliver regular training | $130,780 | 4.5% | 17.7% |
| 5a. Eval employee skills | $116,250 | 4.5% | 15.7% |
| 8a. Codify incident response plan | $114,530 | 4.5% | 15.5% |
| 8b. Test incident response plan | $114,530 | 4.5% | 15.5% |
| 2b. Hunt malicious activity | $100,000 | 4.5% | 13.5% |
| 6b. Test backups | $14,530 | 4.5% | 2.0% |
| 8c. Maintain incident response plan | $14,530 | 4.5% | 2.0% |
| 1a. Deploy MFA | $11,529 | 4.5% | 1.6% |
| 2a. Deploy EDR | $11,529 | 4.5% | 1.6% |
| 3a. Encrypt in transit | $0 | 4.5% | 1.2% |
| 3b. Encrypt at rest | $0 | 4.5% | 1.2% |
| 4a. Remove sharing barriers | $0 | 4.5% | 1.2% |
| 4b. Threat intelligence | $0 | 4.5% | 1.2% |
| 6a. Regular backups | $0 | 4.5% | 1.2% |
| 6c. Protect backups | $0 | 4.5% | 1.2% |
| 6d. Store backups offline | $0 | 4.5% | 1.2% |
| 7a. Timely updates & patching | $0 | 4.5% | 1.2% |
| 7b. Centralized patch system | $0 | 4.5% | 1.2% |
| 7c. Risk-based patching | $0 | 4.5% | 1.2% |
| 9a. External pen testing | $0 | 4.5% | 1.2% |
| 9b. Red team exercises | $0 | 4.5% | 1.2% |
| 10a. Network segmentation | $0 | 4.5% | 1.2% |
| Sums | $628,208 | 100.0% | 100.0% |

Once we have the prorated rates applied to each of the control differences, we sum them together to produce the net weighted security control deviation that is used to compute the Defense Gap Index scalar. This scalar increases or decreases the forecasted impact of incidents based on the organization's security posture. The next step is determining the shape of the loss function relative to security posture.

*Loss/Impact results*

Cybersecurity losses are characterized by non-linear functions with infrequent large losses and a larger number of smaller losses (Eling and Wirfs, 2019). Our results from the benchmarking follow a similar trend with one large loss and three much smaller ones. We take these losses and map them across the maturity levels that are between +/-30% of the average maturity level for the group. We assume that the largest loss would be associated with a security rating near the -30% level, and the small losses at the other end. We also include revealed average loss for the average security posture. One plotted, we estimate a function to capture the relationship (Figure 17). This function is flatter among organizations with better security, but steeper (with losses getting larger faster) as the security level of the organization declines relative to the average. Our fitted exponential equation is shown below allows for losses greater than the USD 500,000 or larger loss that was observed in our data collection. This makes general sense as larger municipal losses have been reported in other states.

Fitted equation using observed loss data:

(Eq 4) $$y = e^{-5.205x}$$

Figure 17 shows the Defense Gap Index scalar in relationship to the net weighted security control deviation. For organizations with the average level of security maturity, the Defense Gap Index scalar is 1 and results forecasts of average losses. However, if the organization's net weighted security is 10% lower than average, the equation above produces a Defense Gap Index scalar of 1.6 which implies that losses will be 1.6 times higher than average when there is an incident.[4] By contrast, when an organization's net weighted security is 10% better than average, we forecast losses to be 41% lower than the average.

**Figure 17: Comparing weighted security control deviation with the Defense Gap Index scalar**

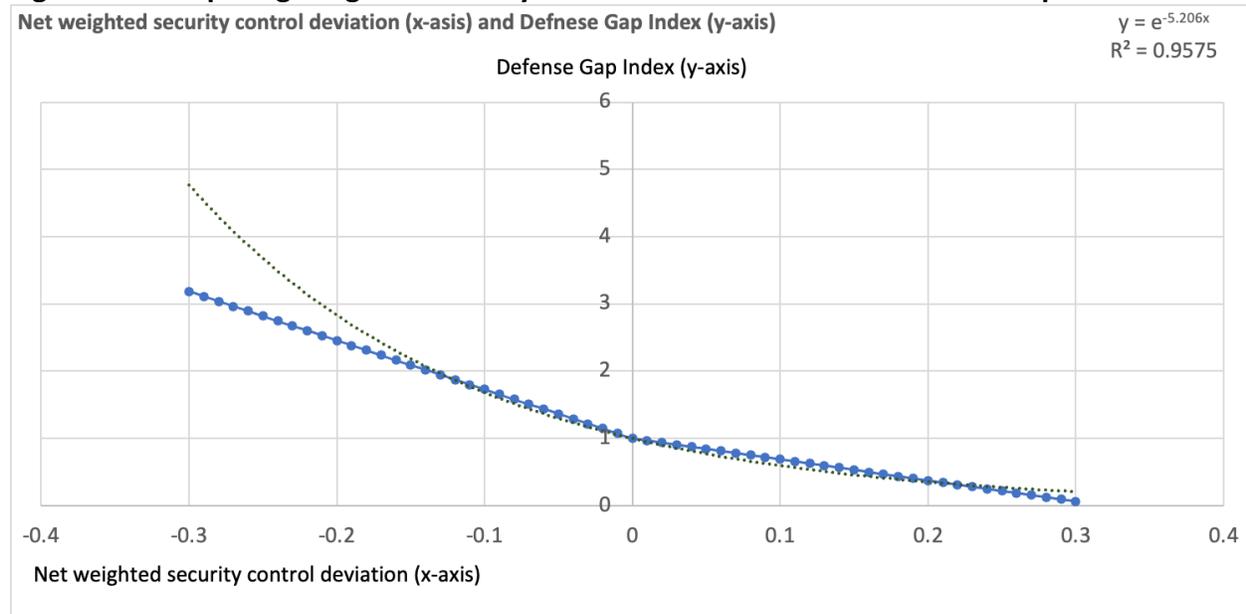

The average reported loss for an incident was USD 157,000, and this is used as the benchmark for an organization with the average level of security maturity.

---

[4] The Defense Gap Index uses a weighted calculation that assigns higher weights to controls exhibiting large losses and lower weights to controls with small or no losses across the group.